\documentclass[a4paper]{jpconf}
\usepackage{graphicx}
\begin{document}
\title{Mode visibilities in radial velocity and intensity Sun-as-a-star helioseismic measurements}

\author{D Salabert$^{1,2}$, J Ballot$^3$ and R A Garc\'ia$^4$}

\address{$^1$ Instituto de Astrof\'isica de Canarias, E-38200 La Laguna, Tenerife, Spain}
\address{$^2$ Departamento de Astrof\'isica, Universidad de La Laguna, E-38206 La Laguna, Tenerife, Spain}
\address{$^3$ Laboratoire d'Astrophysique de Toulouse-Tarbes, Universit\'e de Toulouse, CNRS, F-31400 Toulouse, France}
\address{$^4$ Laboratoire AIM, CEA/DSM-CNRS, Universit\'e Paris 7 Diderot, IRFU/SAp, Centre de Saclay, F-91191 Gif-sur-Yvette, France}

\ead{salabert@iac.es, jerome.ballot@ast.obs-mip.fr, rgarcia@cea.fr}

\begin{abstract}
We analyze more than 5000 days of Sun-as-a-star radial velocity GOLF and intensity VIRGO observations to measure the visibilities of the $l=0$, 1, 2, and 3 modes and the $m$-amplitude ratios of the $l=2$ and 3 modes in the solar acoustic spectrum. We provide observational values that we compare to theoretical predictions.
\end{abstract}

\section{Introduction}
In Sun-as-a-star helioseismology, it is common practice to fix the amplitude ratios between the $m$-components of the $l = 2$ and 3 multiplets (the so-called $m$-amplitude ratio) during the peak-fitting procedure when estimating the p-mode characteristics (e.g., Salabert et al. [1]), while the amplitudes of the $l=1$, 2, and 3 modes relative to the $l=0$ modes are left free (the so-called mode visibility). However, in asteroseismology, the mode visibilities are fixed to theoretical values due to lower signal-to-noise ratio (SNR) and shorter time series, and the $m$-amplitude ratios are expressed as a function of the inclination of the rotation axis only (Appourchaux et al. [2]; Garc\'\i a et al. [3]). In both cases, they are supposed to not depend on the star magnetic activity. However, in the near future, this situation could change when stellar activity cycles will be measured in asteroseismic targets (e.g., Garc\'\i a et al. [4]). After several years of observations collected by the Kepler mission, the SNR will be high enough to measure these parameters in a wide range of solar-like stars in the HR diagram at different evolution stages (Bedding et al. [5]; Chaplin et al. [6]). Moreover, simultaneous observations from SONG (Doppler velocity) and Kepler (intensity) will be extremely useful to better understand stellar atmospheres. 
Although, variations with the height in the solar atmosphere at which the measurements are obtained have been observed in the intensity VIRGO/SPM data at the beginning of the SoHO mission (Fr{\"o}hlich et al. [7]), it has never been verified if these values change in the radial velocity GOLF measurements between the blue- and red-wing observing periods. 

\section{Observations and analysis}
We used observations collected by the space-based instruments Global Oscillations at Low Frequency (GOLF) and Variability of Solar Irradiance and Gravity Oscillations (VIRGO) onboard the {\it Solar and Heliospheric Observatory} (SoHO) spacecraft. GOLF (Gabriel et al. [8]) measures the Doppler velocity at different heights in the solar atmosphere depending on the wing -- Blue or Red -- of the sodium doublet  -- D1 and D2 -- in which the observations were performed (Garc\'\i a et al. [9]). VIRGO (Fr{\"o}hlich et al. [10]) is composed of three Sun photometers (SPM) at 402~nm (blue), 500~nm (green) and 862~nm (red). A total of 5021 days of GOLF and VIRGO observations starting on 1996 April 11 and ending on 2010 January 8 were analyzed, with respective duty cycles of  95.4\% and 94.7\%. The power spectra of the time series were fitted to extract the mode parameters (Salabert et al. [11]) using a standard likelihood maximization function (power spectrum with a $\chi^2$ with 2 d.o.f. statistics). Each mode component was parameterized using an asymmetric Lorentzian profile. Since SoHO observes the Sun equatorwards, only the $l+|m|$ even components are visible in Sun-as-a-star observations of GOLF and VIRGO. In order to obtain observational estimates of the $m$-amplitude ratios, the $m = \pm2$ and $m = 0$ components of the $l = 2$ multiplet, and the $m = \pm3$ and $m = \pm1$ components of the $l = 3$ multiplet were fitted using independent amplitudes, assuming that components with opposite azimuthal order $m$ have the same amplitudes ($H_{l,n,-m}=H_{l,n,+m}$). Note that the blue and red periods of GOLF were also analyzed separately, as well as the mean power spectrum of the three VIRGO  SPMs.

\section{Mode visibilities and $m$-amplitude ratios}
\begin{figure*} 
\begin{center} 
\includegraphics[scale=0.31]{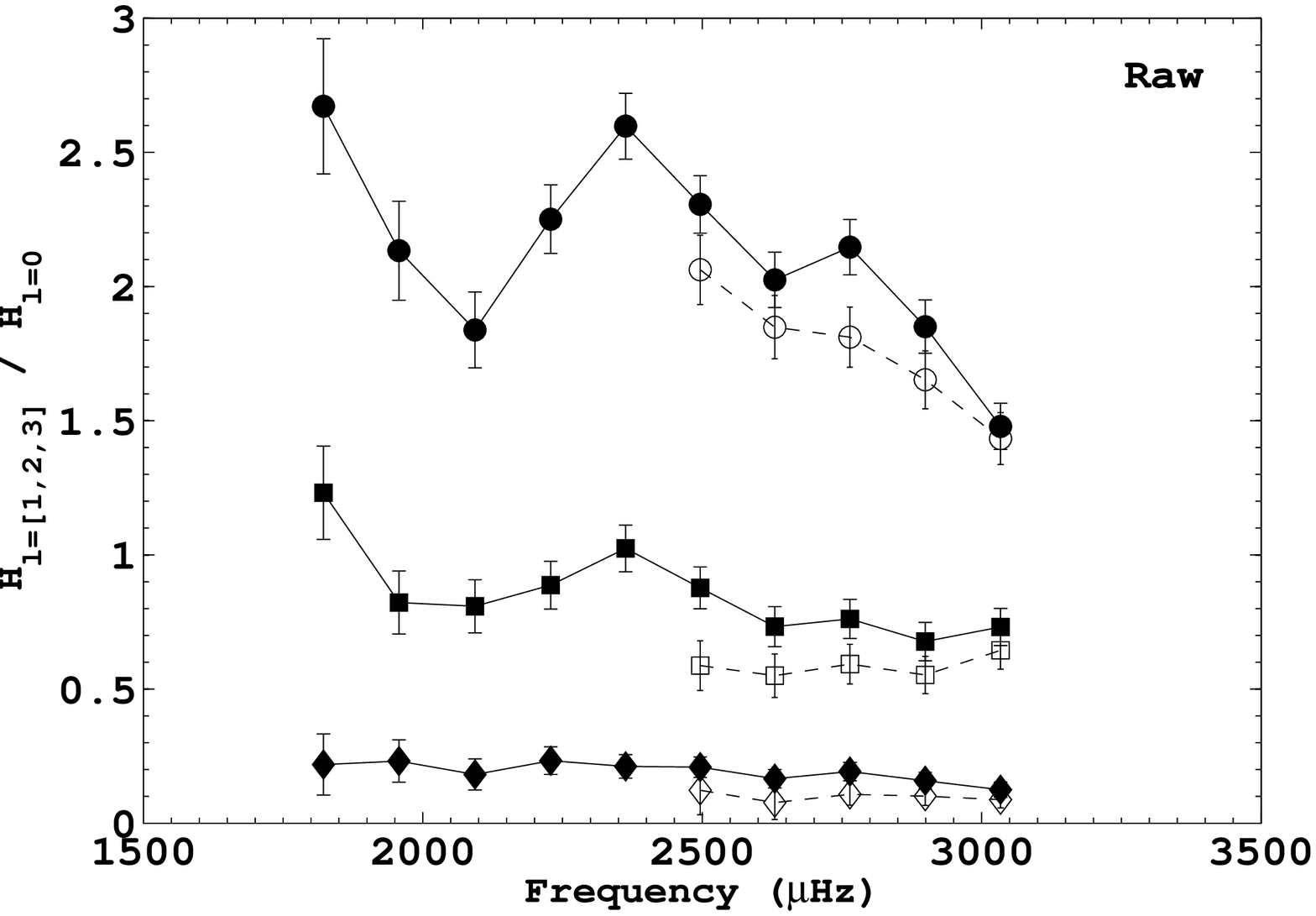}  \includegraphics[scale=0.31]{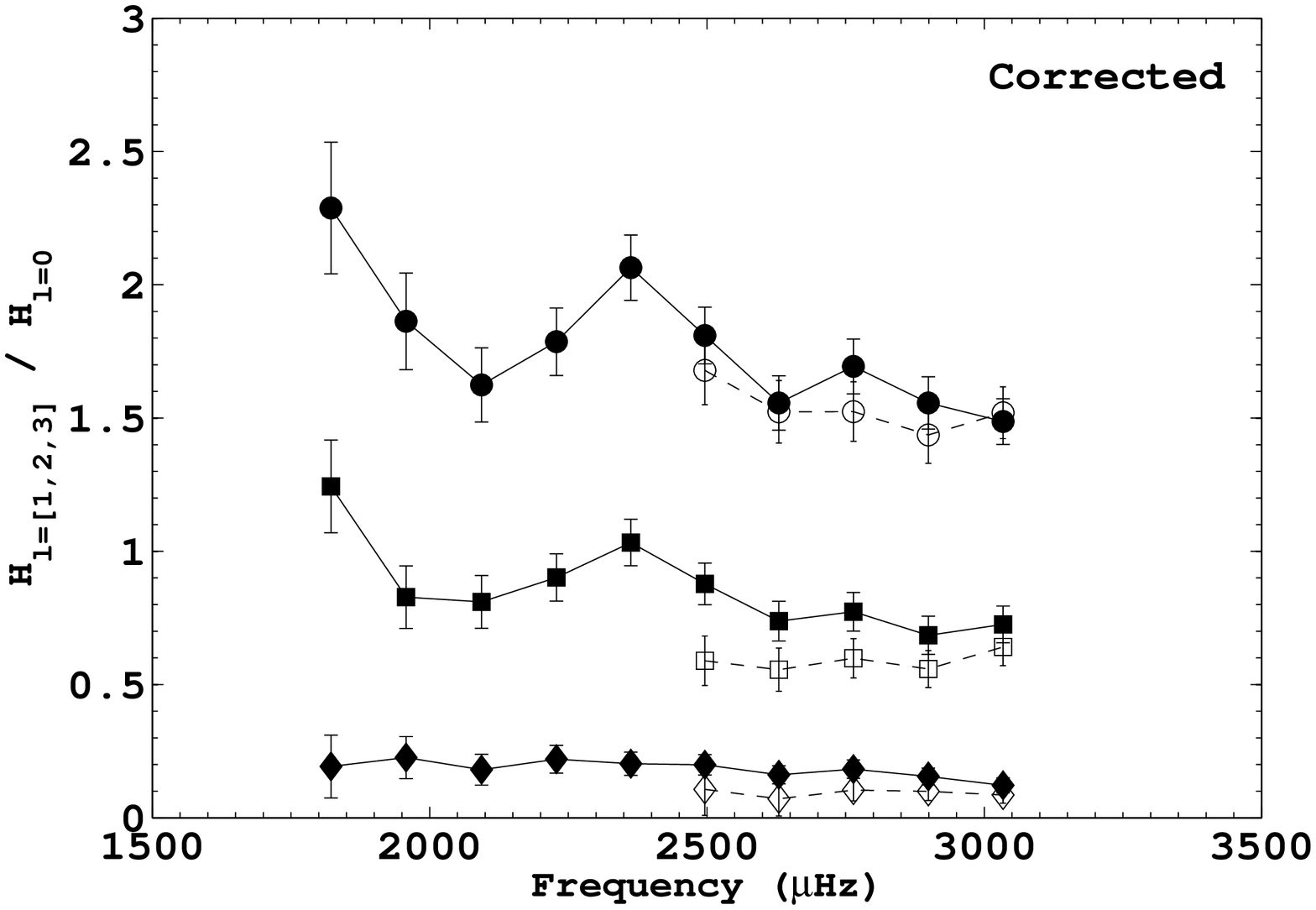}
\end{center} 
\caption{\label{fig:visi} Raw (left) and corrected (right) mode visibilities of $l = 1$ ($\opencircle$), $l=2$ ($\opensquare$), and $l=3$ ($\opendiamond$) relative to $l = 0$ as a function of frequency in GOLF ($\full$) and VIRGO ($\dashed$) observations.} 
\end{figure*}

\begin{figure}
\includegraphics[width=2.5in]{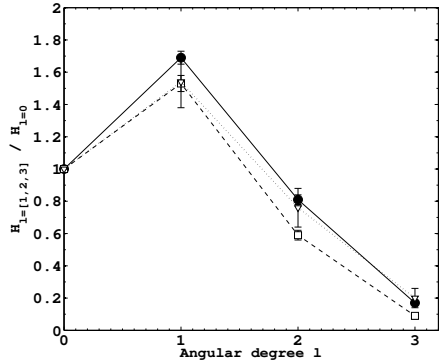}\hspace{0.5pc}%
\begin{minipage}[b]{22pc}\caption{\label{fig:meanvisi} Mode visibilities as a function of angular degree $l$ in GOLF ($\full$, $\fullcircle$) and VIRGO ($\dashed$, $\opensquare$) measurements. For comparison, the mode visibilities of the CoRoT target HD49385 measured by Deheuvels et al. [12] are also represented ($\dotted$, $\opentriangledown$).}
\end{minipage}
\end{figure}

\begin{figure}
\includegraphics[scale=0.25]{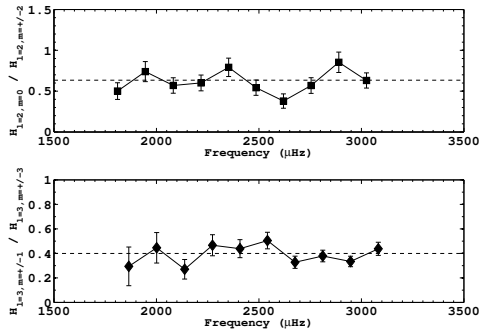}\hspace{0.5pc}%
\begin{minipage}[b]{20.5pc}\caption{\label{fig:mratio} $m$-amplitude ratios of the $l = 2$ (top) and $l = 3$ (bottom) modes as a function of frequency in the GOLF measurements.}
\end{minipage}
\end{figure}

The amplitude of a given multiplet $(l,n)$ is defined as the sum of the amplitudes of its $m$-components, as $H_{l,n} = \sum_{m=-l}^{m=+l} H_{l,n,m}$. 
Then, the visibilities of the $l=1$, 2, and 3 modes relative to the $l=0$ mode are respectively defined as the ratios $H_{l=1,n} / H_{l=0,n}$, $H_{l=2,n-1} / H_{l=0,n}$, and $H_{l=3,n-1}$. The left panel of Fig.~\ref{fig:visi} shows these visibilities for both radial velocity (GOLF) and intensity (VIRGO) measurements as a function of frequency. These raw mode visibilities present a variation with frequency -- especially for the $l=1$ mode -- that is due to the large variation of the mode amplitudes with frequency, even over half a large frequency separation.
Thus, the visibilities are biased and in order to correct them we interpolated (using a spline interpolation) the amplitudes of the $l=1$, 2, and 3 modes to the frequencies of the $l=0$ mode (right panel of Fig.~\ref{fig:visi}). 
Figure~\ref{fig:meanvisi} shows the mode visibilities averaged over frequency for both GOLF and VIRGO observations as a function of $l$  (see Table~\ref{tab:visigolf}). The amplitude ratios between the $m$-components of the $l = 2$ and $l = 3$ multiplets, defined as $H_{l=2,m=0}/H_{l=2,m=\pm2}$ and $H_{l=3,m=\pm1}/H_{l=3,m=\pm3}$ respectively, are represented on Fig.~\ref{fig:mratio} in the case of the radial velocity GOLF measurements and are also reported in Table \ref{tab:mratiogolf}.

\begin{table}
\caption{\label{tab:visigolf} Mode visibilities in radial velocity GOLF and intensity VIRGO measurements.}
\begin{center}
\begin{tabular}{lllll}
\br
Mode visibility & GOLF & GOLF & GOLF& \\
Radial velocity &         & Blue wing & Red wing&\\
\mr
$H_{l=1}/H_{l=0}$ & 1.69$\pm$0.04 & 1.60$\pm$0.05 & 1.85$\pm$0.06&\\
$H_{l=2}/H_{l=0}$ & 0.81$\pm$0.03 & 0.74$\pm$0.04 & 0.98$\pm$0.05&\\
$H_{l=3}/H_{l=0}$ & 0.17$\pm$0.01 & 0.14$\pm$0.02 & 0.28$\pm$0.03&\\
\br
Mode visibility & VIRGO & VIRGO & VIRGO & VIRGO \\
Intensity        &               & Blue      & Green   & Red \\
\mr
$H_{l=1}/H_{l=0}$ & 1.53$\pm$0.05 & 1.55$\pm$0.05 & 1.52$\pm$0.05 & 1.39$\pm$0.05\\
$H_{l=2}/H_{l=0}$ & 0.59$\pm$0.03 & 0.63$\pm$0.03 & 0.57$\pm$0.03 & 0.42$\pm$0.03\\
$H_{l=3}/H_{l=0}$ & 0.09$\pm$0.02 & 0.10$\pm$0.02 & 0.09$\pm$0.02 & 0.05$\pm$0.02\\
\br
\end{tabular}
\end{center}
\end{table}

\begin{table}
\caption{\label{tab:mratiogolf} $m$-amplitude ratios in radial velocity GOLF and intensity VIRGO measurements.}
\begin{center}
\begin{tabular}{lll}
\br
$m$-amplitude ratio & GOLF & VIRGO\\
\mr
$H_{l=2,m=0}/H_{l=2,m=\pm2}$ & 0.63$\pm$0.03 & 0.75$\pm$0.06\\
$H_{l=3,m=\pm1}/H_{l=3,m=\pm3}$ & 0.40$\pm$0.02 & 0.63$\pm$0.06\\
\br
\end{tabular}
\end{center}
\end{table}

\section{Models}
When the contribution of a solar-disk element to the total flux depends only on its distance to the limb, the mode visibility and the $m$-amplitude ratio are decoupled (e.g. Gizon \& Solanki [13]; Ballot et al. [14]). For VIRGO observations, this is verified since the contribution depends mainly on the limb-darkening. However, for GOLF, this is no more the case and we have performed complete computation taking into account the instrumental response, which differ for the blue and red wings. Results of these computations are listed in Table~\ref{tab:visimodel} . The limb-darkening law of Neckel \& Labs [15] has been used. Indeed, visibility values for GOLF vary with frequency by a few percents due to the horizontal motions of modes that increase at low frequency.
In general, these predictions agree with the observations. There is nevertheless some shortcomings: (i) even if the trend is correct, the difference between blue and red wings for GOLF is larger than expected; (ii) the visibility of the $l = 3$ modes in VIRGO are sensitively higher than expected. That could be explained by stronger effects of limb-darkening.

\begin{table}
\caption{\label{tab:visimodel} Modeled visibilities and $m$-amplitude ratios in intensity VIRGO and radial velocity GOLF measurements.}
\begin{center}
\begin{tabular}{llll}
\br
Mode visibility \& & VIRGO & GOLF            & GOLF\\
 $m$-amplitude ratio       &               & Blue  wing   &   Red wing\\
\mr
$H_{l=1}/H_{l=0}$ & 1.51 & 1.84 & 1.86\\
$H_{l=2}/H_{l=0}$ & 0.53 & 1.09 & 1.14\\
$H_{l=3}/H_{l=0}$ & 0.025 & 0.27 & 0.31\\
\mr
$H_{l=2,m=0}/H_{l=2,m=\pm2}$ & 0.67 & 0.59 & 0.58\\
$H_{l=3,m=\pm1}/H_{l=3,m=\pm3}$ & 0.60 & 0.43 & 0.40\\
\br
\end{tabular}
\end{center}
\end{table}

\ack
The authors want to thank Catherine Renaud and Antonio Jim\'enez for the calibration and preparation of the GOLF and VIRGO datasets. The GOLF and VIRGO instruments onboard SoHO are a cooperative effort of many individuals, to whom we are indebted. SoHO is a project of international collaboration between ESA and NASA. DS acknowledges the support of the grant PNAyA2007-62650 from the Spanish National Research Plan. This work has been partially supported by the CNES/GOLF grant at the SAp/CEA-Saclay.

\end{document}